# Misfit Strain Induced Giant Magnetoelectric Coupling in Thin Ferroic Films

M.D. Glinchuk,[1] A.N. Morozovska,[1*] E.A. Eliseev,[1] and R. Blinc[2]

[1]Institute for Problems of Material Science, National Academy of Sciences of Ukraine
Krjijanovskogo 3, 03142 Kiev, Ukraine,

[2]Jožef Stefan Institute, P. O. Box 3000, 1001 Ljubljana, Slovenia

**Abstract**

We show that the misfit strain due to the film-substrate lattice mismatch strongly increases the value of the quadratic magnetoelectric coupling. This giant coupling effect, the size effects and the misfit strain cause strong changes of the phase diagrams of ferroic films at zero external magnetic and electric fields. The antiferromagnetic to ferromagnetic or ferrimagnetic phase transitions for compressive or tensile misfit strains open the way for the tailoring of magnetic and electric properties of ferroic films leading to new applications.

## 1. Introduction

The magnetoelectric (ME) effect is the coupling between the magnetization and polarization, involving different powers of order parameters.[1] The revival of interest in the magnetoelectric effect (ME) is due to numerous possible applications [2,3] as well as to the discovery of relatively high ME effects in both single phase and nanocomposite materials.[4,5,6,7,8,9] The physical reason of the high ME effect is however still unclear in single-phase materials and is the subject of intensive research.[10] Recently we have shown that restricted curved geometry strongly influences the ME coefficients and changes the phase diagrams of ferroic nanorods.[11]

Recently Wang et al. [12] and Tian et al.[13] reported about the dramatically higher ME coefficients and spontaneous polarization values in heteroepitaxially strained thin films of $BiFeO_3$ in comparison with the bulk material. Similar effects are also found in thin polycristaline films.[14] Ruette et al.[15] showed the transition from antiferromagnetic state to ferromagnetic phase

---

[*] Corresponding author: morozo@i.com.ua, permanent address: V. Lashkarev Institute of Semiconductor Physics, NAS of Ukraine, 41, pr. Nauki, 03028 Kiev, Ukraine

order in BiFeO$_3$. The authors assumed that the transition might be induced either by a magnetic field or by epitaxial strain.

In this paper we show that epitaxial misfit strain due to lattice mismatch at the film-substrate interface may significantly change the magnetoelectric coupling coefficients, the surface energy parameters and the polar and magnetic phase diagrams of antiferromagnetic-ferroelectric films. Thus it allows for tailoring of the electric and magnetic properties of ferroic films opening the way to new applications.

**2. Free energy functional**

Let us consider a thin film made of an antiferromagnetic-ferroelectric uniaxial insulating film of thickness $l$ ($l/2 \leq z \leq l/2$) epitaxially grown on a thick rigid substrate. The film is in perfect electric contact with thin planar conducting electrodes [see Fig 1a]. For sake of simplicity we consider that piezomagnetism is absent whereas magnetostriction exists in the bulk of the film.

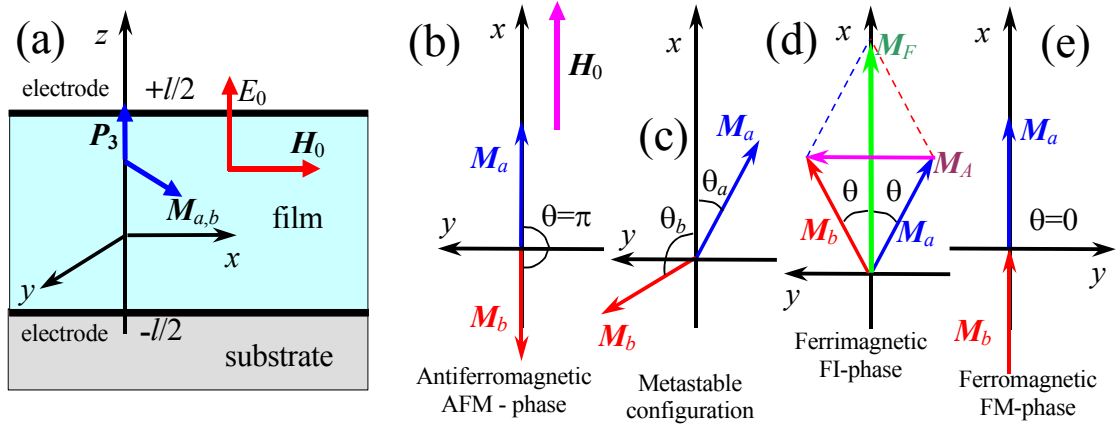

Fig.1. (Color online) (a) Geometry of film: $x$ is the weak magnetic anisotropy axis; $z$ is the polar ferroelectric axis, the external electric field $E_0$ is directed along polar axes, magnetic field $H_0$ is directed along $x$ axes, $\mathbf{M}_a$ and $\mathbf{M}_b$ are sublattices magnetization vectors. (b, d, e) Possible stable magnetic phases: antiferromagnetic phase (AFM), ferrimagnetic phase (FI) and ferromagnetic phase (FM) considered later.

In the Landau-Ginsburg-Devonshire phenomenological theory the free energy is

$$\Delta G = \frac{1}{l}\int_{-l/2}^{l/2} dz\, g_V(z) + G_S\left(\frac{l}{2}\right) + G_S\left(-\frac{l}{2}\right) \tag{1}$$

where $g_V$ and $G_S$ describe order parameter dependent contributions of the bulk and surface of the film. For the description of phase transitions in the ferroelectric-antiferromagnetic films consisting of two magnetic sublattices with magnetization vectors $\mathbf{M}_a$ and $\mathbf{M}_b$, we suppose that polarization $P_3$ and electric field $E_0$ are directed along the polar axis $z$. The axis $x$ is assumed to be the weak magnetic anisotropy axis. When study size-induced phase transitions in thin films the dependence of polarization $P_3$ and magnetization of two sublattices $\mathbf{M}_{a,b}$ on depth $z$ should be considered.[16, 17] The expansion of the Gibbs energy density $g_V$ in terms of the order parameters $P_3$ and $\mathbf{M}_{a,b}$ has the form:

$$g_V = \begin{pmatrix} a_1 P_3^2 + a_{11} P_3^4 + a_{111} P_3^6 + \gamma\left(\dfrac{dP_3}{dz}\right)^2 - Q_{ij33}\sigma_{ij} P_3^2 - \dfrac{A_{ij}}{2}\sigma_{ij}^2 P_3^2 - P_3\left(E_0 + \dfrac{E_d}{2}\right) - \dfrac{s_{ijkl}}{2}\sigma_{ij}\sigma_{kl} \\ b(\mathbf{M}_a^2 + \mathbf{M}_b^2) + c\mathbf{M}_a\mathbf{M}_b + d(\mathbf{M}_a^4 + \mathbf{M}_b^4) + k\mathbf{M}_a^2\mathbf{M}_b^2 + \delta\left(\dfrac{d\mathbf{M}_a}{dz}\right)^2 + \delta\left(\dfrac{d\mathbf{M}_b}{dz}\right)^2 \\ -(M_{a1} + M_{b1})H_0 + b_1(M_{a1}^2 + M_{b1}^2) + c_1 M_{a1} M_{b1} - Z_{ij11}\sigma_{ij}(M_{a1}^2 + M_{b1}^2) - W_{ij11}\sigma_{ij} M_{a1} M_{b1} \end{pmatrix} \quad (2)$$

Subscripts 1, 2 and 3 denote Cartesian coordinates $x$, $y$, $z$ and summation rules are used. We assume that the bulk material has cubic symmetry in the para-phase. The bulk energy, the correlation energy, the interaction with the external field $E_0$, the striction terms, the elastic energy and the depolarization field $E_d$ are included in the Eq.(2). The coefficients $a_1(T) = \alpha_P(T - T_C^b)$ and $b(T) = \alpha_M(T - T_N^b)$ explicitly depend on temperature $T$ whereas all other expansion coefficients are assumed to be temperature independent. Here $T_C^b$ and $T_N^b$ are the Curie and Neel transition temperatures. $\sigma_{ij}$ is elastic stress, $Q_{ijkl}$, $Z_{ijkl}$ and $W_{ijkl}$ are the electro- and magnetostriction coefficients respectively, $s_{ijkl}$ are components of the elastic compliance tensor. Note that the demagnetization field is absent when $M_{a,b3}=0$. Typically $|b| >> |c| >> |b_1| + |c_1|$. Thus we deliberately neglect striction contribution into the highest terms $b$ and $c$, while consider its influence into the weak anisotropy terms. For AFM-phase with weak axis $x$ to be stable in the bulk sample the inequalities $c>0$ and $2b_1 - c_1 < 0$ should be valid (the case $2b_1 - c_1 > 0$ corresponds to the weak plane).

For the case of a single domain insulator film with ideal electrodes the depolarizing field $E_d$ has the form $E_d = 4\pi(\overline{P_3} - P_3(z))$,[17] where the bar designates spatial averaging over the film thickness, i. e. $\overline{P_3} \equiv \int_{-l/2}^{l/2} P_3(z)dz / l$.

The equilibrium equations are obtained after varying the Gibbs energy with respect to the elastic stress $\sigma_{ij}$, $\partial G_V / \partial \sigma_{jk} = -u_{jk}$. The misfit strains $u_{11} = u_{22} = u_m$ are non-zero at the film-substrate boundary $z=-l/2$. The upper surface is free: $\sigma_{3j} = 0$ at $z=l/2$. The nonzero homogeneous stresses are $\sigma_{11} = \dfrac{u_m}{s_{1111} + s_{1122} + A_{11}P_3^2} + \dfrac{u_{22}^S s_{1122} - u_{11}^S(s_{1111} + A_{11}P_3^2)}{(s_{1111} + A_{11}P_3^2)^2 - s_{1122}^2}$ and $\sigma_{22} = \dfrac{u_m}{s_{1111} + s_{1122} + A_{11}P_3^2} + \dfrac{u_{11}^S s_{1122} - u_{22}^S(s_{1111} + A_{11}P_3^2)}{(s_{1111} + A_{11}P_3^2)^2 - s_{1122}^2}$, where the bulk spontaneous strains $u_{11}^S = Q_{1122}P_3^2 + Z_{1111}(M_{a1}^2 + M_{b1}^2) + W_{1111}M_{a1}M_{b1}$ and $u_{22}^S = Q_{1122}P_3^2 + Z_{1122}(M_{a1}^2 + M_{b1}^2) + W_{1122}M_{a1}M_{b1}$. The homogeneous elastic solution is valid until the film thickness is less than the critical thickness $l_d$ for the appearance of misfit dislocation that is known to be dozens of nm. For the film thickness $l > l_d$, an effective misfit strain $u_m^*(l) = u_m l_d / l$ should be introduced in the bulk of the film, while $u_m^*(l) \equiv u_m$ at $l \leq l_d$, $l_d$ being small at high $|u_m|$.[18]

In the vicinity of the surface inversion symmetry breaking takes place and the surface piezoelectric effect $g_{ijk}^e$ has to be taken into account in the surface free energy[19].

$$G_S\left(\pm\frac{l}{2}\right) = \frac{1}{l}\left(\frac{\delta}{\lambda_M}(\mathbf{M}_a^2 + \mathbf{M}_b^2) + \frac{\delta}{\lambda_{MA}}(M_{a1}^2 + M_{b1}^2) + \frac{\gamma}{\lambda_P}P_3^2 - g_{3jk}^e \sigma_{jk} P_3\right)\bigg|_{z=\pm\frac{l}{2}} \quad (3)$$

where $\lambda_p$ and $\lambda_M$, $\lambda_{MA}$ are ferroelectric [17, 20] and magnetic [16] extrapolation lengths respectively, at that $\lambda_{MA} \gg \lambda_M$ allowing for weak magnetic anisotropy.

### 3. Strain effect on phase diagram

Introducing ferromagnetic $\mathbf{M}_F = \mathbf{M}_a + \mathbf{M}_b$ and antiferromagnetic $\mathbf{M}_A = \mathbf{M}_a - \mathbf{M}_b$ order parameters, the condition $\mathbf{M}_a^2 = \mathbf{M}_b^2 = M^2$ is valid allowing for magnetic sublattices equivalence[21] [see also Figs.1b-d]. Substituting the elastic solutions for $\sigma_{ij}$ into the Gibbs energy (1) and making a Legendre transformation, as well as using the direct variational method proposed in Ref.[22], we obtain the Helmholtz free energies of the different phases *DP=AFM, FM, FI* as:[23]

$$F^{DP}[\overline{P}_3, \overline{M}, \theta] \approx \begin{pmatrix} \alpha_P(T - T_{cr}^{FE}(l))\overline{P}_3^2 + a_{11}^m \overline{P}_3^4 - (E_m + E_0)\overline{P}_3 + 16\widetilde{d}\cdot\overline{M}^4 \\ + 2\alpha_M(T - T_{cr}^{DP}(l))\overline{M}^2 + 4\widetilde{f}^{DP}(l)\overline{P}_3^2\overline{M}^2 + \Delta F^{DP}[\overline{M}, \theta] \end{pmatrix}, \quad (4a)$$

$$\Delta F^{AFM} = 0, \quad \Delta F^{FM} = -2H_0\overline{M}, \quad \Delta F^{FI} \approx (2c + 4\widetilde{c}_1)\overline{M}^2\overline{\cos\theta}^2 - 2H_0\overline{M}\cdot\overline{\cos\theta}. \quad (4b)$$

The expressions for the renormalized coefficients in Eqs.(4) are summarized in Table I. In the AFM-phase the non-zero component is $M_{A1}(z) \equiv 2M(z)$ [see Fig.1b], while $M_{F1}(z) \equiv 2M(z)$ is non-zero in FM-phase [see Fig.1e]. Here we explicitly take the depth $z$ dependence of the order parameters into account. In the FI-phase $\mathbf{M}_F(z) = (2M(z)\cos\theta(z),0,0)$ and $\mathbf{M}_A(z) = (0,2M(z)\sin\theta(z),0)$ are [see Fig.1d and Ref.[24], [25]], and $\overline{\cos\theta} \approx \left(c + 2\tilde{c}_1 + 2\tilde{f}^{FM}\overline{P_3^2}\right)^{-1} H_0 / 2\overline{M}$ for a single-domain case and high extrapolation length $\lambda_{MA}$. The asymmetric phase (c) is unstable in the bulk at arbitrary magnetic field. The film phase (c) is unstable at zero magnetic field. At zero magnetic field $H_0=0$ the angle $\theta = \pi/2$ and the absolute stability of the FI-phase corresponds to the weak axes - weak plane phase transition.

**Table I.** Free energy renormalized coefficients

| Temperatures | Expansion and ME coefficients |
|---|---|
| $T_{cr}^{FE}(l) \approx T_C^b + \dfrac{2Q_{12}u_m^*(l)}{\alpha_P(s_{11}+s_{12})} - \dfrac{2\gamma}{\alpha_P l\left(\sqrt{\gamma/4\pi}+\lambda_P\right)}$ | $a_{11}^m(l) = a_{11} + \dfrac{Q_{12}^2}{s_{11}+s_{12}} + \dfrac{2A_{11}Q_{12}u_m^*(l)}{(s_{11}+s_{12})^2}$ |
| $T_{cr}^{FI}(l) \approx T_N^b + \dfrac{c}{2\alpha_M} - \dfrac{4\pi^2\delta}{\alpha_M l(\pi^2\lambda_M + 2l)}$ | $\dfrac{\tilde{f}^{AFM}(l)}{f_-} = \dfrac{\tilde{f}^{FM}(l)}{f_+} = 1 + \dfrac{u_m^*(l)A_{11}}{(s_{11}+s_{12})Q_{12}}$ |
| $T_{cr}^{AFM}(l) \approx T_N^b + \dfrac{c}{2\alpha_M} - \dfrac{4\pi^2\delta}{\alpha_M l(\pi^2\lambda_{eff} + 2l)} - \dfrac{2\tilde{b}_1}{\alpha_M}$ | $f_\pm = Q_{12}\dfrac{2(Z_{11}+Z_{12}) \pm (W_{11}+W_{12})}{4(s_{11}+s_{12})}$ |
| $T_{cr}^{FM}(l) \approx T_N^b - \dfrac{c}{2\alpha_M} - \dfrac{4\pi^2\delta}{\alpha_M l(\pi^2\lambda_{eff} + 2l)} - \dfrac{2\tilde{c}_1}{\alpha_M}$ | $\tilde{f}^{FI}(l,\theta) \equiv \tilde{f}^{FM}(l)\overline{\cos\theta}^2$, $\tilde{d} = \dfrac{d}{8} + \dfrac{k}{16}$ |
| $\tilde{b}_1(l) = \dfrac{b_1}{2} - \dfrac{c_1}{4} - \dfrac{f_-}{Q_{12}}u_m^*(l)$, $\tilde{c}_1(l) = \dfrac{b_1}{2} + \dfrac{c_1}{4} - \dfrac{f_+}{Q_{12}}u_m^*(l)$ | $\lambda_{eff} = \dfrac{\lambda_M \lambda_{MA}}{\lambda_M + \lambda_{MA}}$, $E_m(l) = \dfrac{2u_m g_{311}^e}{s_{11}+s_{12}} \cdot \dfrac{\lambda_P}{l\gamma}$ |
|  | (Voigt's notations are used). |

The averaged magnetization $\overline{M}$ depends on polarization $\overline{P_3}$ via the magnetoelectric coupling $\tilde{f}^{DP}$ by the following way:

$$\overline{M}^2 = -\left(\alpha_M\left(T - T_{cr}^{DF}(l)\right) + 2\tilde{f}^{DP}\overline{P_3^2}\right)/16\tilde{d}. \quad (5)$$

So, coupling induced phase transitions could appear. At zero resulting field $E_m + E_0 = 0$ each of the phases (4) could be either paraelectric (PE) at $P_3=0$ or ferroelectric (FE) at $P_3\neq 0$. Estimation of material parameters shows that size effects and misfit strain substantially renormalize the free energy coefficients. The misfit strain may significantly increase the values of the quadratic magnetoelectric coupling coefficients $\tilde{f}^{AFM,FM}(l)$ in comparison with bulk values $f_\pm$.

The dependence of the normalized magnetoelectric coupling coefficients $\tilde{f}^{AFM}/f_-$ and $\tilde{f}^{FM}/f_+$ on the film thickness for different misfit strains is illustrated in Fig 2a. Because $\tilde{f}^{AFM,FM}$ can be positive or negative, they lead to an increase or a decrease of the order parameters as shown in Fig. 2b for polarization $P_3 \neq 0$.

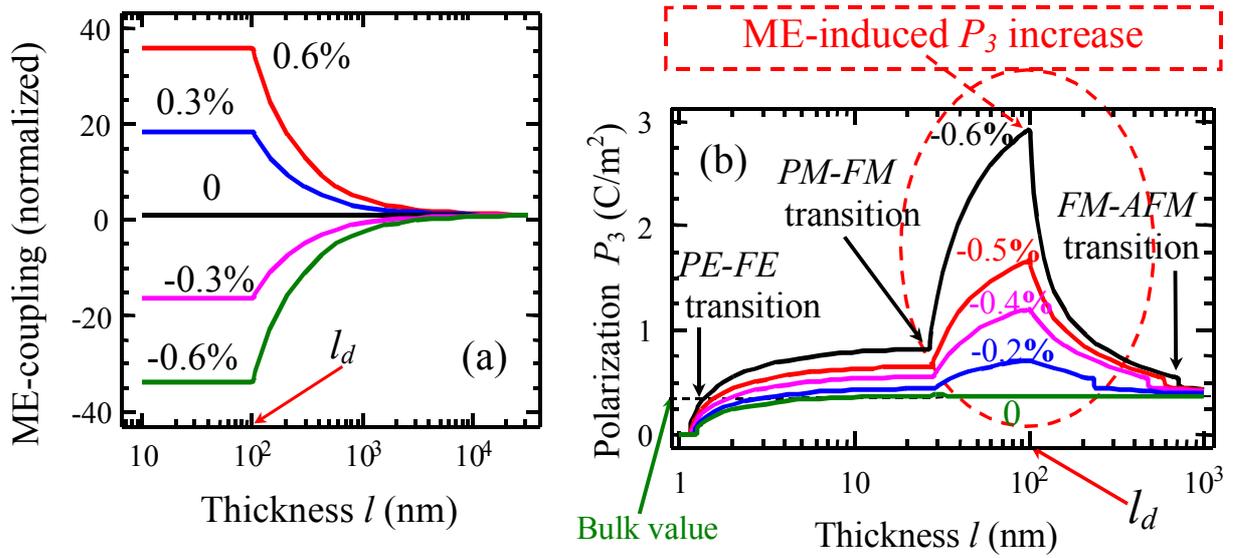

Fig.2. (Color online) (a) Normalized ME-coupling coefficients $\tilde{f}^{AFM}/f_- = \tilde{f}^{FM}/f_+$ and (b) polarization $P_3$ dependence on film thickness $l$ for different misfit strain $u_m$ in % (labels near the curves) and $l_d$=100 nm. Reasonable material parameters in SI units: $a_1(T)=(T-1103)\cdot 5\cdot 10^5$, $a_{11}=6.5\cdot 10^8$, $b(T)=(T-642)\cdot 10^{-5}$, $T$=300 K; $\gamma=10^{-9}$, $\delta=10^{-20}$, $c=10^{-5}$, $b_1=-5\cdot 10^{-6}$, $c_1=10^{-7}$, $a_{11}=6.5\cdot 10^8$, $d=10^{-15}$, $k=3\cdot 10^{-16}$, $s_{11}=5.3\cdot 10^{-12}$, $s_{12}=-1.85\cdot 10^{-12}$, $Q_{12}=-0.005$, $Z_{11}=W_{11}=-10^{-14}$, $Z_{12}=W_{12}=4\cdot 10^{-15}$, $A_{11}=-10^{-10}$; $g_{31}^e=0$. Lengths $\lambda_P$=4nm, $\lambda_M$=0.4 nm, $\lambda_{MA}$=400nm.

It should be stressed that the order parameters $\overline{M}=\overline{M}(T,l,u_m)$ and $\overline{P}_3=\overline{P}_3(T,l,u_m)$ can be tuned by the misfit strain $u_m$ and film thickness $l$, thus leading to size- and ME coupling-induced phase transitions. The significant increase of the polarization compared to the bulk is clearly seen from Fig.2b.

Let us now show the changes of phase diagrams and the possibility of the appearance of the ferromagnetic phase appearance at zero external magnetic and electric fields (i.e. $H_0$=0 and $E_0$=0). The phase diagrams of strained ferroic films at zero external fields are shown in Figs.3 for reasonable material parameters. The stabilization of the AFM phase with the increase of the sublattice interaction constant $c$ is similar to what is known for bulk materials [compare Fig.3a with 3b,c]. It is clear that size effects and misfit strain (at film thickness less than the critical thickness for the appearance of the misfit dislocations $l_d$) cause strong changes of phase diagrams. In particular, ferromagnetic and ferrimagnetic phases may appear in thin film antiferromagnetic bulk. Small extrapolation length and depolarization field effects decrease the corresponding order parameter value and cause thickness-induced paraelectric phase transition in thin ferroic films [compare Figs.3a-c for $\lambda_P$=4 nm with Figs.3d-e for $\lambda_P$=0.4 nm]. For $\lambda_P$=4 nm a paraelectric phase transition appears at film thicknesses <0.3nm (not shown). The relatively high magnetic extrapolation length $\lambda_M$=4nm is responsible for the steep boundaries between several magnetic phases in Figs. 3f. The decrease of $\lambda_P$ or $\lambda_M$ stabilizes paraelectric or paramagnetic phases respectively. This is so because the extrapolation length reflects the rate of polarization or magnetization profile change with film thickness, so that the thinner the film the sharper is the decrease of polarization or magnetization profile. This increases the region of $P_3$=0 or $M$=0, i.e. the region of the existence of PE or PM phases.

The nonzero surface piezoeffect coefficient $g_{31}^e$ immediately leads to the appearance of nonzero built-in electric field $E_m \neq 0$ that induces polarization $P_3 \neq 0$ and thus turns the paraelectric phase into the electret-like one [see Fig.3f with *E*-phases instead of *PE*-phases]. Since $E_m \sim 1/l$, the induced polarization shifts the phase boundaries at small thickness $l$ allowing for the quadratic magnetoelectric coupling.

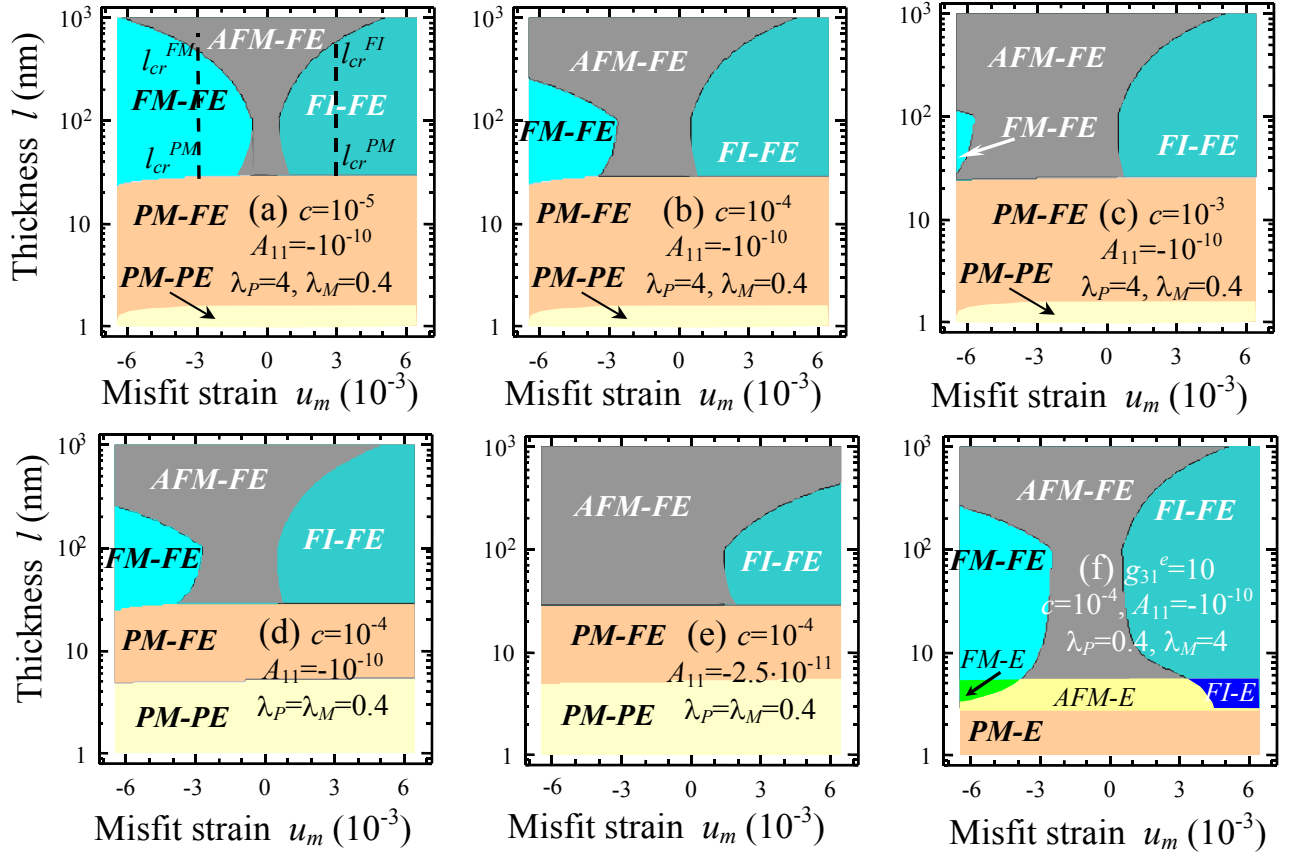

Fig. 3. (Color online). Phase diagrams of strained ferroic films: *AFM-FE* designates an antiferromagnetic and ferroelectric phase, *FM-FE* is ferromagnetic and ferroelectric phase (secondary ferroic phase), *FI-FE* is ferrimagnetic (week plane at $\theta = \pi/2$) and ferroelectric phase, *PM-FE* is paramagnetic-ferroelectric phase, *PM-PE* is paramagnetic-paraelectric phase. External fields are zero. Built-in electric field is absent ($g_{31}^e=0$) for plots (a)-(e), while $g_{31}^e=10$ for plot (f). Letter *E* designates electret-like phase. Material parameters are listed in Fig.2. Different values of $A_{11}$ and $c$, $\lambda_P$ and $\lambda_M$ (in nm) are listed in plot labels.

It is seen that the transformation of the ferroic film phase diagrams from the antiferromagnetic to ferromagnetic or ferrimagnetic once under compressive or tensile strains is a general feature of ferroic films. This phenomenon can be observed for the film thickness $l_{cr}^{PM} < l < l_{cr}^{FM}$ and $l_{cr}^{PM} < l < l_{cr}^{FI}$, where $l_{cr}^{FM}$, $l_{cr}^{FI}$ and $l_{cr}^{PM}$ are respectively critical thickness of the AFM–FM, AFM–FI and FM–PM phase transitions. The values of $l_{cr}^{FM}$ and $l_{cr}^{FI}$ depend on the misfit strain value (see dashed vertical lines in Fig.3a for $u_m = -0.3\%$ and $u_m = 0.3\%$), while $l_{cr}^{PM}$ is defined by the almost horizontal boundary between magnetic and paramagnetic phases, indicating that the critical thickness appeared independent of the misfit strain. Actually $l_{cr}^{PM}(T) \approx (4\delta/\lambda_M \alpha_M)(T_N^b - T + c/2\alpha_M)^{-1}$.

At a small value of the striction coupling constant $|A_{11}|$ the AFM→FM transition disappears [compare Figs.3c and 3d]. The bulk ME coupling terms $f_{\pm}$ are typically small. Since the product $A_{11}u_m^*$ is absent in the bulk, $A_{11}u_m^*$ supports the appearance of the ferromagnetism FM-FE in thin films.

**Summary**

The size-induced antiferromagnetic to ferromagnetic phase transition in thin films of ferroics is most probable for compressive misfit strains more than $10^{-3}$, negative electrostriction coefficient $Q_{12}$, relatively high striction coupling constant $|A_{11}|$ and small sublattices interaction constant $c$. In contrast to the appearance ferromagnetic phase transition for compressive strains, tensile misfit strain about $10^{-3}$ or higher may cause antiferromagnetic spin-flop transition from the weak anisotropy axis into perpendicular plane at zero external magnetic field, i.e. a spontaneous size-induced weak axis- weak plane transition.

The predicted increase of polarization, giant magnetoelectric coupling and the appearance of ferromagnetism in thin antiferromagnetic films are in qualitative agreement with available experimental data [9, 12, 13, 15]. In BiFeO$_3$ ferroelectric-antiferromagnetic thin films of thickness 70-400 nm on SrTiO$_3$ substrate [12] the corresponding compressive misfit strain $u_m$ varies from -0.5% up to -1% depending on the film growth temperature. Estimations on the basis of the free energy (5) with reasonable material parameters[26] and $l_d \sim$ 10-100 nm lead to the ferromagnetic phase stability in BiFeO$_3$ in the thicknesses range $20 \le l \le 500$ nm. It appeared that polarization in BiFeO$_3$ thin films is essentially higher than in the bulk material.[9, 12, 13] Our calculations showed that the increase for 5-10 times is caused by giant ME coupling [see Fig. 2b]. A more rigorous comparison is hardly possible, since bismuth ferrite can be regarded uniaxial antiferromagnetic only approximately (spins in the neighboring atoms are antiparallel) and the majority of its electric and magnetic parameters are not measured.

It is worth to stress that practically the same strong increase of the ME coupling could be obtained in strained ferromagnetic-ferroelectric films. In the polydomain case the inhomogeneous strain in the vicinity of thin ferroelectric domain walls via strong electrostriction may cause a local ferromagnetic phase transition in antiferromagnetic ferroelectrics.

The obtained results open the way for tailoring the magnetic and electric properties of ferroic films leading to new applications.

**Acknowledgement**

This work is supported in part (MDG and RB) by MULTICERAL. ANM gratefully acknowledges financial support from National Academy of Science of Ukraine and Ministry of Science and Education of Ukraine.

*Supplementary materials to the manuscript*

**Misfit Strain Induced Giant Magnetoelectric Coupling in Thin Ferroic Films**

M.D. Glinchuk[1], A.N. Morozovska[1*], E.A. Eliseev[1], and R. Blinc[2]

[1]Institute for Problems of Material Science, NAS of Ukraine

Krjijanovskogo 3, 03142 Kiev, Ukraine,

[2]Jožef Stefan Institute, P. O. Box 3000, 1001 Ljubljana, Slovenia

**Appendix A**

Gibbs energy expansion on the order parameters $P_3(T,l,z)$, $\mathbf{M}_{a,b}(T,l,z)$ has the form:

$$G_V = \frac{1}{l}\int_{-l/2}^{l/2} dz \left(\begin{array}{l} a_1(T)P_3^2 + a_{11}P_3^4 + a_{111}P_3^6 - (Q_{11}\sigma_3 + Q_{12}(\sigma_1+\sigma_2))P_3^2 + \gamma\left(\frac{dP_3}{dz}\right)^2 - \\ -\frac{A_{11}}{2}(\sigma_1^2+\sigma_2^2+\sigma_3^2)P_3^2 - P_3\left(E_0 + \frac{1}{2}E_d\right) + \\ + b(T)(\mathbf{M}_a^2+\mathbf{M}_b^2) + c\mathbf{M}_a\mathbf{M}_b + d(\mathbf{M}_a^4+\mathbf{M}_b^4) + k\mathbf{M}_a^2\mathbf{M}_b^2 + \\ + b_1(M_{a1}^2+M_{b1}^2) + c_1 M_{a1}M_{b1} - (M_{a1}+M_{b1})H_0 - \\ -(Z_{11}\sigma_1 + Z_{12}(\sigma_2+\sigma_3))(M_{a1}^2+M_{b1}^2) + \delta\left[\left(\frac{d\mathbf{M}_a}{dz}\right)^2 + \left(\frac{d\mathbf{M}_b}{dz}\right)^2\right] - \\ -(W_{11}\sigma_1 + W_{12}(\sigma_2+\sigma_3))M_{a1}M_{b1} - \frac{1}{2}s_{11}(\sigma_1^2+\sigma_2^2+\sigma_3^2) \\ -s_{12}(\sigma_1\sigma_2+\sigma_1\sigma_3+\sigma_3\sigma_2) - \frac{1}{2}s_{44}(\sigma_4^2+\sigma_5^2+\sigma_6^2) \end{array}\right) \quad \text{(A.1a)}$$

Subscripts 1, 2 and 3 denote Cartesian coordinates *x*, *y*, *z* and Voigt's notations are used. We assume that bulk material paraelectric phase has cubic symmetry. The bulk energy, the correlation energy, the interaction with the external field $E_0$, striction terms, elastic energy and the depolarization field $E_d$ are included in the expansion (2). The coefficients $a_1(T) = \alpha_P(T - T_C^b)$ and $b(T) = \alpha_M(T - T_N^b)$ explicitly depend on temperature whereas all other expansion coefficients are assumed to be temperature independent. Here $T_C^b$ and $T_N^b$ are the Curie and Neel transition temperatures. $\sigma_i$ is elastic stress $Q_{ij}$, $Z_{ij}$ and $W_{ij}$ are the electro- and magnetostriction coefficients respectively whereas $s_{ij}$ are components of the elastic compliance tensor. Note that demagnetization field is absent when $M_{a,b3}=0$. For the sake of simplicity we omitted the terms like $B_{11}\sigma_{ij}^2 M_{a,bj}^2$, since $B_{11}$ values are regarded small and unknown for magnetics, in contrast to the known values $A_{ii} \approx A_{11}$ for ferroelectrics. Typically $|b| \gg |c| \gg |b_1| + |c_1|$. Thus we deliberately neglect striction contribution into the highest terms $b$ and $c$, while consider its influence into the weak anisotropy terms. For AFM-phase with weak axis $x$ to be stable in the bulk sample the inequalities $c>0$ and $2b_1 - c_1 < 0$ should be valid (the case $2b_1 - c_1 > 0$ corresponds to the weak plane).

---

* Corresponding author: morozo@i.com.ua, permanent address: V. Lashkarev Institute of Semiconductor Physics, NAS of Ukraine, 41, pr. Nauki, 03028 Kiev, Ukraine

Substituting elastic solution into Eqs.(A.1) and making Legendre transformations from the Gibbs energy $G$ to the Helmholtz free energy $F$, we obtained:

$$F_V = \frac{1}{l}\int_{-l/2}^{l/2} dz \begin{cases} a_1^m(T)P_3^2 + a_{11}^m P_3^4 + \gamma\left(\dfrac{dP_3}{dz}\right)^2 - P_3\left(E_0 + \dfrac{1}{2}E_d\right) + \\ + b(T)(\mathbf{M}_a^2 + \mathbf{M}_b^2) + c(T)\mathbf{M}_a\mathbf{M}_b + d(\mathbf{M}_a^4 + \mathbf{M}_b^4) + k\mathbf{M}_a^2\mathbf{M}_b^2 + \\ + b_1^m(M_{a1}^2 + M_{b1}^2) + c_1^m M_{a1}M_{b1} + \delta\left[\left(\dfrac{d\mathbf{M}_a}{dz}\right)^2 + \left(\dfrac{d\mathbf{M}_b}{dz}\right)^2\right] \\ + g_{11}P_3^2(M_{a1}^2 + M_{b1}^2) + f_{11}P_3^2 M_{a1}M_{b1} - (M_{a1} + M_{b1})H_0 \end{cases} \quad \text{(A.1b)}$$

Where renormalized coefficients are introduced as:

$$a_1^m(T,l) = a_1(T) - \frac{2Q_{12}u_m^*(l)}{s_{11} + s_{12}}, \quad \text{(A.2a)}$$

$$b_1^m = b_1 - \frac{Z_{11} + Z_{12}}{s_{11} + s_{12}} u_m^*(l), \quad \text{(A.2b)}$$

$$c_1^m(l) = c_1 - \frac{W_{11} + W_{12}}{s_{11} + s_{12}} u_m^*(l) \quad \text{(A.2c)}$$

$$a_{11}^m(l) = a_{11} + \frac{Q_{12}^2}{s_{11} + s_{12}} + \frac{2A_{11}Q_{12}u_m^*(l)}{(s_{11} + s_{12})^2}, \quad \text{(A.2d)}$$

ME coupling coefficients:

$$f_{11}(l) = \frac{(W_{11} + W_{12})}{s_{11} + s_{12}}\left(Q_{12} + \frac{u_m^*(l)A_{11}}{s_{11} + s_{12}}\right), \quad \text{(A.2e)}$$

$$g_{11}(l) = \frac{(Z_{11} + Z_{12})}{s_{11} + s_{12}}\left(Q_{12} + \frac{u_m^*(l)A_{11}}{s_{11} + s_{12}}\right). \quad \text{(A.2f)}$$

Introducing conventional ferromagnetic $\mathbf{M}_F = \mathbf{M}_a + \mathbf{M}_b$ and antiferromagnetic $\mathbf{M}_A = \mathbf{M}_a - \mathbf{M}_b$ order parameters (and thus $\mathbf{M}_a = (\mathbf{M}_F + \mathbf{M}_A)/2$ and $\mathbf{M}_b = (\mathbf{M}_F - \mathbf{M}_A)/2$), one rewrite bulk free energy (A.1) as:

$$F_V = \frac{1}{l}\int_{-l/2}^{l/2} dz \begin{pmatrix} a_1^m P_3^2 + a_{11}^m P_3^4 + \gamma\left(\frac{dP_3}{dz}\right)^2 - P_3\left(E_0 + \frac{1}{2}E_d\right) + \\ + \tilde{c}\mathbf{M}_F^2 + \tilde{b}\mathbf{M}_A^2 + \tilde{d}\left(\mathbf{M}_F^2 + \mathbf{M}_A^2\right)^2 + \\ + \frac{\delta}{2}\left(\frac{d\mathbf{M}_F}{dz}\right)^2 + \frac{\delta}{2}\left(\frac{d\mathbf{M}_A}{dz}\right)^2 + \\ + \tilde{c}_1 M_{F1}^2 + \tilde{b}_1 M_{A1}^2 + \tilde{g}_{11} P_3^2 M_{F1}^2 + \tilde{f}_{11} P_3^2 M_{A1}^2 - M_{F1} H_0 \end{pmatrix} \quad (A.3)$$

Here we used that $(\mathbf{M}_F \mathbf{M}_A) = 0$, since the absolute value $\mathbf{M}_a^2 = \mathbf{M}_b^2 = M^2(T,l,z)$ [see Fig.1b-d].

The coefficients in Eq.(A.3) are expressed via the coefficients (A.2) as $\tilde{b} = \frac{b}{2} - \frac{c}{4}$, $\tilde{c} = \frac{b}{2} + \frac{c}{4}$, $\tilde{b}_1 = \frac{b_1^m}{2} - \frac{c_1^m}{4}$, $\tilde{c}_1 = \frac{b_1^m}{2} + \frac{c_1^m}{4}$, $\tilde{d} = \frac{d}{8} + \frac{k}{16}$, $\tilde{g}_{11} = \frac{g_{11}}{2} + \frac{f_{11}}{4}$, $\tilde{f}_{11} = \frac{g_{11}}{2} - \frac{f_{11}}{4}$. For AFM-phase to be stable in the bulk the inequality $\tilde{b} < \tilde{c}$ should be valid, which is true for $c>0$.

Surface free energy (3) acquires the form:

$$F_S \approx \frac{1}{l}\begin{pmatrix} \frac{\delta}{\lambda_M}\left[\mathbf{M}_F^2\left(\frac{l}{2}\right) + \mathbf{M}_F^2\left(-\frac{l}{2}\right) + \mathbf{M}_A^2\left(\frac{l}{2}\right) + \mathbf{M}_A^2\left(-\frac{l}{2}\right)\right] + \\ \frac{\delta}{\lambda_{MA}}\left[M_{F1}^2\left(\frac{l}{2}\right) + M_{F1}^2\left(-\frac{l}{2}\right) + M_{A1}^2\left(\frac{l}{2}\right) + M_{A1}^2\left(-\frac{l}{2}\right)\right] + \\ \frac{\gamma}{\lambda_P}\left(P_3^2\left(\frac{l}{2}\right) + P_3^2\left(-\frac{l}{2}\right)\right) - \frac{2g_{311}^e u_m}{s_{11}+s_{12}}\left(P_3\left(\frac{l}{2}\right) + P_3\left(-\frac{l}{2}\right)\right) \end{pmatrix} \quad (A.4)$$

(I). AFM-phase. In AFM-phase ferromagnetic order parameter is absent, while antiferromagnetic one is pointed along easy axis "1". In this case the component $M_{A1} \equiv 2M(z)$ is non-zero, so that free energy (A.3)-(A.4) reduces to:

$$F_V^{AFM}[P_3, M] = \frac{1}{l}\int_{-l/2}^{l/2} dz \begin{pmatrix} a_1^m P_3^2 + a_{11}^m P_3^4 + \gamma\left(\frac{dP_3}{dz}\right)^2 - P_3\left(E_0 + \frac{1}{2}E_d\right) + \\ 4\left(\tilde{b} + \tilde{b}_1 + \tilde{f}_{11}P_3^2\right)M^2 + 16\tilde{d}\cdot M^4 + 2\delta\left(\frac{dM}{dz}\right)^2 \end{pmatrix} \quad (A.5a)$$

$$F_S^{AFM}[P_3, M] \approx \begin{pmatrix} \frac{4\delta}{l}\left(\frac{1}{\lambda_M} + \frac{1}{\lambda_{MA}}\right)\left(M^2\left(\frac{l}{2}\right) + M^2\left(-\frac{l}{2}\right)\right) + \\ \frac{\gamma}{\lambda_P l}\left(P_3^2\left(\frac{l}{2}\right) + P_3^2\left(-\frac{l}{2}\right)\right) - \frac{2g_{311}^e u_m}{(s_{11}+s_{12})l}\left(P_3\left(\frac{l}{2}\right) + P_3\left(-\frac{l}{2}\right)\right) \end{pmatrix} \quad (A.5b)$$

Variation of the free energy (A.5) leads to the Euler-Lagrange equations:

$$2\left(a_1^m + 4\widetilde{f}_{11}M^2\right)P_3 + 4a_{11}^m P_3^3 - 2\gamma\frac{d^2 P_3}{dz^2} = E_0 + E_d, \quad (A.6a)$$

$$\left(\widetilde{b} + \widetilde{b}_1 + \widetilde{f}_{11}P_3^2\right)M + 8\widetilde{d}\,M^3 - \frac{\delta}{2}\frac{d^2 M}{dz^2} = 0, \quad (A.6b)$$

with boundary conditions:

$$\left(P_3 - \lambda_P \frac{dP_3}{dz}\right)\bigg|_{z=-l/2} = \frac{g_{311}^e u_m}{s_{11} + s_{12}}, \quad \left(P_3 + \lambda_P \frac{dP_3}{dz}\right)\bigg|_{z=l/2} = \frac{g_{311}^e u_m}{s_{11} + s_{12}}, \quad (A.6c)$$

$$\left(\frac{M}{\lambda_{eff}} - \left(\frac{dM}{dz}\right)\right)\bigg|_{z=-l/2} = 0, \quad \left(\frac{M}{\lambda_{eff}} + \left(\frac{dM}{dz}\right)\right)\bigg|_{z=l/2} = 0 \quad (A.6d)$$

Hereinafter $\lambda_{eff} = \frac{\lambda_M \lambda_{MA}}{\lambda_M + \lambda_{MA}}$. Then using direct variational method proposed in Ref.[22] with trial functions for polarization $P_3$ listed there and for magnetization $M$ as for in-plane polarization components $P_{1,2}$ listed there we obtained free energy Eqs.(4) with renormalized coefficients from Eqs.(A.5). Minimization of the free energy (4) leads to the coupled equations in AF-phase:

$$\begin{cases} 2\left(\alpha_P\left(T - T_{cr}^{FE}(l)\right) + 4\widetilde{f}_{11}\overline{M}^2\right)\overline{P}_3 + 4a_{11}^m \overline{P}_3^3 = E_m + E_0 \\ \left(\widetilde{b}(T) + \widetilde{b}_1 + \widetilde{f}_{11}\overline{P}_3^2 + \frac{2\pi^2\delta}{l\left(\pi^2 \lambda_{eff} + 2l\right)}\right)\overline{M} + 8\widetilde{d}\cdot\overline{M}^3 = 0 \end{cases}, \quad (A.7a)$$

It is easy to obtain that non-zero solution $\overline{M}^2 = -\frac{1}{8\widetilde{d}}\left(\widetilde{b}(T) + \widetilde{b}_1 + \frac{2\pi^2\delta}{l\left(\pi^2 \lambda_{eff} + 2l\right)} + \widetilde{f}_{11}\overline{P}_3^2\right)$ exists under the condition $\left(\widetilde{b}(T) + \widetilde{b}_1 + \frac{2\pi^2\delta}{l\left(\pi^2 \lambda_{eff} + 2l\right)} + \widetilde{f}_{11}\overline{P}_3^2\right) < 0$, since $\widetilde{d} > 0$. Corresponding free energy

$$F^{AFM}\left[\overline{P}_3\right] \approx \left(\begin{array}{l}\left(\alpha_P\left(T - T_{cr}^{FE}(l)\right) - \frac{\widetilde{f}_{11}}{2\widetilde{d}}\left(\widetilde{b}(T) + \widetilde{b}_1 + \frac{2\pi^2\delta}{l\left(\pi^2 \lambda_{eff} + 2l\right)}\right)\right)\overline{P}_3^2 \\ + \left(a_{11}^m - \frac{\widetilde{f}_{11}^2}{4\widetilde{d}}\right)\overline{P}_3^4 - (E_m + E_0)\overline{P}_3 - \frac{1}{4\widetilde{d}}\left(\widetilde{b}(T) + \widetilde{b}_1 + \frac{2\pi^2\delta}{l\left(\pi^2 \lambda_{eff} + 2l\right)}\right)^2\end{array}\right). \quad (A.7b)$$

In particular case of zero electric field $E_m + E_0 = 0$, one obtains polarization value

$$\overline{P}_3^2 = -\frac{\alpha_P\left(T - T_{cr}^{FE}(l)\right) - \left(\widetilde{b}(T) + \widetilde{b}_1 + 2\pi^2\delta/l\left(\pi^2 \lambda_{eff} + 2l\right)\right)\widetilde{f}_{11}/2\widetilde{d}}{2a_{11}^m - \widetilde{f}_{11}^2/2\widetilde{d}}$$ and corresponding free energy

$$F^{AFM} \approx -\frac{\left(\alpha_P(T-T_{cr}^{FE}(l))-\left(\tilde{b}(T)+\tilde{b}_1+2\pi^2\delta/l(\pi^2\lambda_{eff}+2l)\right)\tilde{f}_{11}/2\tilde{d}\right)^2}{4a_{11}^m-\tilde{f}_{11}^2/\tilde{d}}-\frac{1}{4\tilde{d}}\left(\tilde{b}(T)+\tilde{b}_1+\frac{2\pi^2\delta}{l(\pi^2\lambda_{eff}+2l)}\right)^2$$

(II). <u>FM-phase.</u> In FM-phase the antiferromagnetic order parameter is absent (see e.g. Refs. [25]), while the ferromagnetic one is pointed along external magnetic field direction. In this case only the component $M_{F1}\equiv 2M(z)$ is non-zero, so free energy (A.3)-(A.4) reduces to:

$$F_V^{FM}[P_3,M]=\frac{1}{l}\int_{-l/2}^{l/2}dz\left(\begin{array}{c}a_1^m P_3^2+a_{11}^m P_3^4+\gamma\left(\frac{dP_3}{dz}\right)^2-P_3\left(E_0+\frac{1}{2}E_d\right)+\\ +4(\tilde{c}+\tilde{c}_1+\tilde{g}_{11}P_3^2)M^2+16\tilde{d}\cdot M^4+2\delta\left(\frac{dM}{dz}\right)^2-2MH_0\end{array}\right) \quad (A.8a)$$

$$F_S^{FM}[P_3,M]\approx\left(\begin{array}{c}4\frac{\delta}{l}\left(\frac{1}{\lambda_M}+\frac{1}{\lambda_{MA}}\right)\left(M^2\left(\frac{l}{2}\right)+M^2\left(-\frac{l}{2}\right)\right)+\\ \frac{\gamma}{\lambda_P l}\left(P_3^2\left(\frac{l}{2}\right)+P_3^2\left(-\frac{l}{2}\right)\right)-\frac{2g_{311}^e u_m}{(s_{11}+s_{12})l}\left(P_3\left(\frac{l}{2}\right)+P_3\left(-\frac{l}{2}\right)\right)\end{array}\right) \quad (A.8b)$$

Using direct variational method in Eqs.(A.8) we obtained corresponding Eq.(4). Minimization of the free energy (4) leads to the coupled equation in FM-phase:

$$\begin{cases}2(\alpha_P(T-T_{cr}^{FE}(l))+4\tilde{g}_{11}\overline{M}^2)\overline{P}_3+4a_{11}^m\overline{P}_3^3\approx E_m+E_0\\ \left(\tilde{c}(T)+\tilde{c}_1+\tilde{g}_{11}\overline{P}_3^2+\frac{2\pi^2\delta}{l(\pi^2\lambda_{eff}+2l)}\right)\overline{M}+8\tilde{d}\cdot\overline{M}^3=\frac{1}{4}H_0\end{cases} \quad (A.9)$$

At $H_0=0$, the non-zero solution $\overline{M}^2=-\frac{1}{8\tilde{d}}\left(\tilde{c}(T)+\tilde{c}_1+\frac{2\pi^2\delta}{l(\pi^2\lambda_{eff}+2l)}+\tilde{g}_{11}\overline{P}_3^2\right)$ exists under the condition $\left(\tilde{c}(T)+\tilde{c}_1+\frac{2\pi^2\delta}{l(\pi^2\lambda_{eff}+2l)}+\tilde{g}_{11}\overline{P}_3^2\right)<0$, since $\tilde{d}>0$. Corresponding free energy

$$F^{FM}[\overline{P}_3]\approx\left(\begin{array}{c}\left(\alpha_P(T-T_{cr}^{FE}(l))-\frac{\tilde{g}_{11}}{2\tilde{d}}\left(\tilde{c}(T)+\tilde{c}_1+\frac{2\pi^2\delta}{l(\pi^2\lambda_{eff}+2l)}\right)\right)\overline{P}_3^2+\left(a_{11}^m-\frac{\tilde{g}_{11}^2}{4\tilde{d}}\right)\overline{P}_3^4\\ -(E_m+E_0)\overline{P}_3-\frac{1}{4\tilde{d}}\left(\tilde{c}(T)+\tilde{c}_1+\frac{2\pi^2\delta}{l(\pi^2\lambda_{eff}+2l)}\right)^2\end{array}\right). \quad (A.10)$$

In particular case of zero electric field $E_m + E_0 = 0$, one obtains polarization

$$\overline{P}_3^2 = -\frac{\alpha_P(T - T_{cr}^{FE}(l)) - (\tilde{c}(T) + \tilde{c}_1 + 2\pi^2\delta/l(\pi^2\lambda_{eff} + 2l))\tilde{g}_{11}/2\tilde{d}}{2a_{11}^m - \tilde{g}_{11}^2/2\tilde{d}}$$

and free energy

$$F^{FM} \approx -\frac{(\alpha_P(T - T_{cr}^{FE}(l)) - (\tilde{c}(T) + \tilde{c}_1 + 2\pi^2\delta/l(\pi^2\lambda_{eff} + 2l))\tilde{g}_{11}/2\tilde{d})^2}{4a_{11}^m - \tilde{g}_{11}^2/\tilde{d}} - \frac{1}{4\tilde{d}}\left(\tilde{c}(T) + \tilde{c}_1 + \frac{2\pi^2\delta}{l(\pi^2\lambda_{eff} + 2l)}\right)^2.$$

(III). FI-phase. During the possible transition from AFM-phase to FM-phase representations $\mathbf{M}_a = (M\cos\theta, M\sin\theta, 0)$ and $\mathbf{M}_b = (M\cos\theta, -M\sin\theta, 0)$ always may be chosen from symmetry considerations and appropriate coordinate system rotation in mixed ferromagnetic (FI) phase. This immediately leads to the expressions $\mathbf{M}_F(z) = (2M(z)\cos(\theta(z)), 0, 0)$ and $\mathbf{M}_A(z) = (0, 2M(z)\sin\theta(z), 0)$. Thus, in the new variables $\{P_3(z,T,l), \theta(T,z,l)\}$ free energy (A.3)-(A.4) acquires the form

$$F_V^{FI}[P_3, M, \theta] = \frac{1}{l}\int_{-l/2}^{l/2} dz \left( \begin{array}{c} a_1^m P_3^2 + a_{11}^m P_3^4 + \gamma\left(\frac{dP_3}{dz}\right)^2 - P_3\left(E_0 + \frac{1}{2}E_d\right) + \\ + 4(\tilde{c} + \tilde{c}_1 - \tilde{b} + \tilde{g}_{11}P_3^2)M^2\cos^2\theta + 4\tilde{b}M^2 + 16\tilde{d}M^4 + \\ + 2M^2\delta\left(\frac{d\theta}{dz}\right)^2 + 2\delta\left(\frac{dM}{dz}\right)^2 - 2M\cdot H_0\cos\theta \end{array} \right) \quad \text{(A.11a)}$$

$$F_S^{FI}[P_3, M, \theta] \approx \left( \begin{array}{c} \frac{4\delta}{\lambda_M l}\left(M^2\left(\frac{l}{2}\right) + M^2\left(-\frac{l}{2}\right)\right) + \\ + \frac{4\delta}{\lambda_{MA}l}\left(M^2\left(\frac{l}{2}\right)\cos^2\left(\theta\left(\frac{l}{2}\right)\right) + M^2\left(-\frac{l}{2}\right)\cos^2\left(\theta\left(-\frac{l}{2}\right)\right)\right) + \\ \frac{\gamma}{\lambda_P l}\left(P_3^2\left(\frac{l}{2}\right) + P_3^2\left(-\frac{l}{2}\right)\right) - \frac{2g_{311}^e u_m}{(s_{11} + s_{12})l}\left(P_3\left(\frac{l}{2}\right) + P_3\left(-\frac{l}{2}\right)\right) \end{array} \right) \quad \text{(A.11b)}$$

Euler-Lagrange equations acquires the form of nonlinear coupled system:

$$2(a_1^m + 4\tilde{g}_{11}\cos^2\theta M^2)P_3 + 4a_{11}^m P_3^3 - 2\gamma\frac{d^2P_3}{dz^2} = E_0 + E_d, \quad \text{(A.12a)}$$

$$2M\sin 2\theta(\tilde{c} + \tilde{c}_1 - \tilde{b} + \tilde{g}_{11}P_3^2) + 2\delta M\frac{d^2\theta}{dz^2} + 4\delta M\frac{dM}{dz}\frac{d\theta}{dz} = H_0\sin\theta, \quad \text{(A.12b)}$$

$$(\tilde{c} + \tilde{c}_1 - \tilde{b} + \tilde{g}_{11}P_3^2)M\cos^2\theta + \tilde{b}M + 8\tilde{d}M^3 + \frac{\delta}{2}\left(\frac{d\theta}{dz}\right)^2 M - \frac{\delta}{2}\frac{d^2M}{dz^2} = \frac{H_0}{4}\cos\theta. \quad \text{(A.12c)}$$

Boundary conditions:

$$\left(P_3 - \lambda_P \frac{dP_3}{dz}\right)\bigg|_{z=-l/2} = \frac{g^e_{311} u_m}{s_{11}+s_{12}}, \quad \left(P_3 + \lambda_P \frac{dP_3}{dz}\right)\bigg|_{z=l/2} = \frac{g^e_{311} u_m}{s_{11}+s_{12}}, \quad \text{(A.13a)}$$

$$\left(\frac{\sin 2\theta}{\lambda_{MA}} + \left(\frac{d\theta}{dz}\right)\right)\bigg|_{z=-l/2} = 0, \quad \left(\frac{\sin 2\theta}{\lambda_{MA}} - \left(\frac{d\theta}{dz}\right)\right)\bigg|_{z=l/2} = 0 \quad \text{(A.13b)}$$

$$\left(M\left(\frac{1}{\lambda_M} + \frac{\cos^2\theta}{\lambda_{MA}}\right) - \left(\frac{dM}{dz}\right)\right)\bigg|_{z=-l/2} = 0, \quad \left(M\left(\frac{1}{\lambda_M} + \frac{\cos^2\theta}{\lambda_{MA}}\right) + \left(\frac{dM}{dz}\right)\right)\bigg|_{z=l/2} = 0 \quad \text{(A.13c)}$$

(III.a) <u>Zero magnetic field.</u> Single-domain solution $\theta \equiv \pi/2$ (and so $d\theta/dz = 0$) of Eqs.(A.12b) and (A.13b) always exists in the case $H_0$=0 (spin-flop from weak axis $x$ to axis $y$ in the weak plain $yz$; possible transition to $z$-axes is suppressed by demagnetization field). Substitution $\theta = \pi/2$ and $d\theta/dz = 0$ into Eqs.(A.12c) and (A.13c) at $H_0$=0 leads to

$$\tilde{b} M + 8\tilde{d} M^3 - \frac{\delta}{2}\frac{d^2 M}{dz^2} = 0. \quad \text{(A.14a)}$$

$$\left(\frac{M}{\lambda_M} - \left(\frac{dM}{dz}\right)\right)\bigg|_{z=-l/2} = 0, \quad \left(\frac{M}{\lambda_M} + \left(\frac{dM}{dz}\right)\right)\bigg|_{z=l/2} = 0 \quad \text{(A.14b)}$$

After substitution of Eq.(A.14) into the free energy (A.11) we obtained:

$$F_V^{WP}[P_3, M, \theta = \pi/2] = \frac{1}{l}\int_{-l/2}^{l/2} dz \begin{pmatrix} a_1^m P_3^2 + a_{11}^m P_3^4 + \gamma\left(\frac{dP_3}{dz}\right)^2 - P_3\left(E_0 + \frac{1}{2}E_d\right) + \\ + 4\tilde{b} M^2 + 16\tilde{d} M^4 + 2\delta\left(\frac{dM}{dz}\right)^2 \end{pmatrix}, \quad \text{(A.15a)}$$

$$F_S^{WP}[P_3, M, \theta = \pi/2] \approx \begin{pmatrix} \frac{4\delta}{\lambda_M l}\left(M^2\left(\frac{l}{2}\right) + M^2\left(-\frac{l}{2}\right)\right) + \\ \frac{\gamma}{\lambda_P l}\left(P_3^2\left(\frac{l}{2}\right) + P_3^2\left(-\frac{l}{2}\right)\right) - \frac{2g^e_{311} u_m}{(s_{11}+s_{12})l}\left(P_3\left(\frac{l}{2}\right) + P_3\left(-\frac{l}{2}\right)\right) \end{pmatrix} \quad \text{(A.15b)}$$

Using direct variational method we obtained Eqs.(5b) from Eqs.(A.15). Minimization of the free energy (5b) leads to the coupled equation in FI-phase in zero magnetic field:

$$\begin{cases} 2\alpha_P\left(T - T_{cr}^{FE}(l)\right)\overline{P_3} + 4a_{11}^m \overline{P_3}^3 \approx E_m + E_0 \\ \left(\tilde{b}(T) + \frac{2\pi^2 \delta}{l(\pi^2 \lambda_M + 2l)}\right)\overline{M} + 8\tilde{d}\cdot \overline{M}^3 = 0 \end{cases} \quad \text{(A.16)}$$

It is easy to obtain that non-zero solution $\overline{M}^2 = -\dfrac{1}{8\tilde{d}}\left(\tilde{b}(T) + \dfrac{2\pi^2\delta}{l(\pi^2\lambda_M + 2l)}\right)$ exists under the condition $\left(\tilde{b}(T) + \dfrac{2\pi^2\delta}{l(\pi^2\lambda_M + 2l)}\right) < 0$, since $\tilde{d} > 0$. Corresponding free energy

$$F^{FI}[\overline{P}_3] \approx \alpha_P(T - T_{cr}^{FE}(l))\overline{P}_3^2 + a_{11}^m \overline{P}_3^4 - (E_m + E_0)\overline{P}_3 - \dfrac{1}{4\tilde{d}}\left(\tilde{b}(T) + \dfrac{2\pi^2\delta}{l(\pi^2\lambda_M + 2l)}\right)^2. \quad (A.17)$$

In particular case $E_m + E_0 = 0$, one obtains $\overline{P}_3^2 = -\dfrac{\alpha_P(T - T_{cr}^{FE}(l))}{2a_{11}^m}$ and

$$F^{FI} \approx -\dfrac{(\alpha_P(T - T_{cr}^{FE}(l)))^2}{4a_{11}^m} - \dfrac{1}{4\tilde{d}}\left(\tilde{b}(T) + \dfrac{2\pi^2\delta}{l(\pi^2\lambda_M + 2l)}\right)^2 \quad \text{at } H_0=0 \text{ and } \theta = \pi/2.$$

(III.b) Zero magnetic field. Poly-domain solutions may satisfy the conditions $d\theta/dz \neq 0$ under $H_0=0$. Under the assumption $P_3^2 \to \overline{P}_3^2$, $M^2 \to \overline{M}^2$ from Eq.(A.12b) we obtained the expression for the first integral for angle $\theta$ as

$$\dfrac{d\theta}{dz} = \pm\sqrt{\dfrac{1}{\delta}\left(2\cos^2\theta(\tilde{c} + \tilde{c}_1 - \tilde{b} + \tilde{g}_{11}\overline{P}_3^2) + C_0\right)} \quad (A.18)$$

Eq.(A.18) along with boundary condition (A.13b) leads to the expression for the integration constant $C_0 = -2(\tilde{c} + \tilde{c}_1 - \tilde{b} + \tilde{g}_{11}\overline{P}_3^2)\cos^2\left(\theta\left(\pm\dfrac{l}{2}\right)\right) + \delta\left(\dfrac{1}{\lambda_{MA}}\sin\left(2\theta\left(\pm\dfrac{l}{2}\right)\right)\right)^2$. Then equation (A.16) for $\cos(\theta(z))$ acquires the form

$$\dfrac{d\theta}{\sqrt{(\cos^2\theta + C_0/2(\tilde{c} + \tilde{c}_1 - \tilde{b} + \tilde{g}_{11}\overline{P}_3^2))}} = \pm dz\sqrt{\dfrac{2(\tilde{c} + \tilde{c}_1 - \tilde{b} + \tilde{g}_{11}\overline{P}_3^2)}{\delta}} \quad (A.19)$$

For $\mu = \cos(\theta(z))$ we obtained the elliptic integral $\dfrac{d\mu}{\sqrt{(1-\mu^2)(\mu^2 + a)}} = \mp dzb$ leading to $\mu(z) = sn(\sqrt{ab}(z - z_0), -1/a)$, or

$$\cos(\theta(z)) = sn\left(\dfrac{w(z - z_0)}{\sqrt{m}}, m\right), \quad (A.20)$$

Here $z_0$ is integration constant, dimensionless parameter $m = -\frac{2}{C_0}\left(\tilde{c} + \tilde{c}_1 - \tilde{b} + \tilde{g}_{11}\overline{P}_3^2\right)$ and characteristic width $w = \sqrt{-\frac{2}{\delta}\left(\tilde{c} + \tilde{c}_1 - \tilde{b} + \tilde{g}_{11}\overline{P}_3^2\right)}$. Constants $m$ and $z_0$ should be found from the boundary condition (A.13b), namely

$$\mathrm{sn}\left(\frac{w(z-z_0)}{\sqrt{m}}, m\right)\left(\frac{2}{\lambda_{MA}}\mathrm{cn}\left(\frac{w(z-z_0)}{\sqrt{m}}, m\right) + \mathrm{dn}\left(\frac{w(z-z_0)}{\sqrt{m}}, m\right)\frac{w}{\sqrt{m}}\right)\bigg|_{z=-l/2} = 0. \quad \text{(A.21a)}$$

$$\mathrm{sn}\left(\frac{w(z-z_0)}{\sqrt{m}}, m\right)\left(\frac{2}{\lambda_{MA}}\mathrm{cn}\left(\frac{w(z-z_0)}{\sqrt{m}}, m\right) - \mathrm{dn}\left(\frac{w(z-z_0)}{\sqrt{m}}, m\right)\frac{w}{\sqrt{m}}\right)\bigg|_{z=+l/2} = 0. \quad \text{(A.21b)}$$

Then one can use the identity $\mathrm{dn} = \sqrt{1 - m \cdot \mathrm{sn}^2}$. After substitution of Eq.(A.20) into the free energy (A.11a) we obtained:

$$F_V^{FI}[P_3, M] = \frac{1}{l}\int_{-l/2}^{l/2} dz \begin{pmatrix} a_1^m P_3^2 + a_{11}^m P_3^4 + \gamma\left(\frac{dP_3}{dz}\right)^2 - P_3\left(E_0 + \frac{1}{2}E_d\right) + 16\tilde{d} M^4 + \\ + 4\tilde{b}M^2 + 4M^2\left(\tilde{c} + \tilde{c}_1 - \tilde{b} + \tilde{g}_{11}P_3^2\right)\left(2\mathrm{sn}^2(z) - \mathrm{sn}^2\left(\frac{l}{2}\right)\right) \\ + 8M^2\frac{\delta}{\lambda_{MA}^2}\mathrm{sn}^2\left(\frac{l}{2}\right)\mathrm{cn}^2\left(\frac{l}{2}\right) + 2\delta\left(\frac{dM}{dz}\right)^2 \end{pmatrix} \quad \text{(A.22a)}$$

Where we used that $C_0 = -2\left(\tilde{c} + \tilde{c}_1 - \tilde{b} + \tilde{g}_{11}\overline{P}_3^2\right)\mathrm{sn}^2\left(\frac{l}{2}\right) + \frac{4\delta}{\lambda_{MA}^2}\mathrm{sn}^2\left(\frac{l}{2}\right)\mathrm{cn}^2\left(\frac{l}{2}\right)$. Surface energy (A.11b) acquires the form

$$F_S^{FI}[P_3, M] \approx \begin{pmatrix} \frac{4\delta}{\lambda_M l}\left(M^2\left(\frac{l}{2}\right) + M^2\left(-\frac{l}{2}\right)\right) + \\ \frac{4\delta}{\lambda_{MA} l}\left(M^2\left(\frac{l}{2}\right)\mathrm{sn}^2\left(\frac{l}{2}\right) + M^2\left(-\frac{l}{2}\right)\mathrm{sn}^2\left(-\frac{l}{2}\right)\right) + \\ \frac{\gamma}{\lambda_P l}\left(P_3^2\left(\frac{l}{2}\right) + P_3^2\left(-\frac{l}{2}\right)\right) - \frac{2g_{311}^e u_m}{(s_{11}+s_{12})l}\left(P_3\left(\frac{l}{2}\right) + P_3\left(-\frac{l}{2}\right)\right) \end{pmatrix} \quad \text{(A.22b)}$$

Comparing the free energies (A.22) with (A.15) we obtained that (A.15) is lower for positive extrapolation length $\lambda_{MA}$, since coefficient $\left(\tilde{c} + \tilde{c}_1 - \tilde{b} + \tilde{g}_{11}P_3^2\right) \approx \frac{c}{2} + \tilde{g}_{11}P_3^2$ is positive (typically $c > 2\left|\tilde{g}_{11}P_3^2\right|$) and $\mathrm{sn}^2(z) \geq \mathrm{sn}^2(\pm l/2)$ for single-domain case.

(III.c) Non-zero magnetic field ($H_0 \neq 0$). Under the assumption $P_3^2 \rightarrow \overline{P}_3^2$, $M \rightarrow \overline{M}$ from Eq.(A.12b) we obtained the expression for the first integral for angle $\theta$ as

$$\frac{d\theta}{dz} = \pm\sqrt{\frac{1}{\delta}\left(2\cos^2\theta\left(\tilde{c} + \tilde{c}_1 - \tilde{b} + \tilde{g}_{11}\overline{P}_3^2\right) - \frac{H_0}{\overline{M}}\cos\theta + C_0\right)} \qquad (A.23)$$

Eq.(A.23) along with boundary condition (A.13b) leads to the expression for the integration constant $C_0 = \frac{H_0}{\overline{M}}\cos\left(\theta\left(\pm\frac{l}{2}\right)\right) - 2\left(\tilde{c} + \tilde{c}_1 - \tilde{b} + \tilde{g}_{11}\overline{P}_3^2\right)\cos^2\left(\theta\left(\pm\frac{l}{2}\right)\right) + \delta\left(\frac{1}{\lambda_{MA}}\sin\left(2\theta\left(\pm\frac{l}{2}\right)\right)\right)^2$.

Under the conditions for all $\lambda_i \rightarrow \infty$, for angle $\theta$ we obtained expression:

$$\cos\theta = \frac{H_0}{4M\left(\tilde{c} + \tilde{c}_1 - \tilde{b} + \tilde{g}_{11}\overline{P}_3^2\right)}. \qquad (A.24)$$

IV Asymmetric configurations instability at zero magnetic field

Let us consider general representation $\mathbf{M}_a = (M\cos\theta_a, M\sin\theta_a, 0)$ and $\mathbf{M}_b = (M\cos\theta_b, -M\sin\theta_b, 0)$. Substituting the expressions into Eq.(2), making Legendre transformations from the Gibbs energy $G$ to the Helmholtz free energy $F$, we obtained:

$$F_V = \frac{1}{l}\int_{-l/2}^{l/2} dz \begin{pmatrix} a_1^m(T)P_3^2 + a_{11}^m P_3^4 + \gamma\left(\frac{dP_3}{dz}\right)^2 - P_3\left(E_0 + \frac{1}{2}E_d\right) + \\ + 2b(T)M^2 + c(T)M^2\cos(\theta_a + \theta_b) + (2d+k)M^4 + \\ + b_1^m M^2\left(\cos(\theta_a)^2 + \cos(\theta_b)^2\right) + c_1^m M^2\cos(\theta_a)\cos(\theta_b) + \\ + \delta\left[2\left(\frac{dM}{dz}\right)^2 + M^2\left(\frac{d\theta_a}{dz}\right)^2 + M^2\left(\frac{d\theta_b}{dz}\right)^2\right] + \\ + g_{11}P_3^2 M^2\left(\cos(\theta_a)^2 + \cos(\theta_b)^2\right) + f_{11}P_3^2 M^2\cos(\theta_a)\cos(\theta_b) \\ - (\cos(\theta_a) + \cos(\theta_b))M H_0 \end{pmatrix} \qquad (A.25)$$

Equations of state for the angles $\theta_a$ and $\theta_b$ acquire the form:

$$\begin{pmatrix} -c(T)M^2\sin(\theta_a + \theta_b) - \left(b_1^m M^2 + g_{11}P_3^2 M^2\right)2\cos(\theta_a)\sin(\theta_a) \\ -\left(c_1^m M^2 + f_{11}P_3^2 M^2\right)\sin(\theta_a)\cos(\theta_b) - \delta\frac{d}{dz}\left[2M^2\left(\frac{d\theta_a}{dz}\right)\right] - M H_0 \end{pmatrix} = 0 \qquad (A.26a)$$

$$\begin{pmatrix} -c(T)M^2\sin(\theta_a + \theta_b) - \left(b_1^m M^2 + g_{11}P_3^2 M^2\right)2\cos(\theta_b)\sin(\theta_b) \\ -\left(c_1^m M^2 + f_{11}P_3^2 M^2\right)\sin(\theta_b)\cos(\theta_a) - \delta\frac{d}{dz}\left[2M^2\left(\frac{d\theta_b}{dz}\right)\right] - M H_0 \end{pmatrix} = 0 \qquad (A.26b)$$

Introducing the difference and sum of the angles $(\theta_a - \theta_b)$ and $(\theta_a + \theta_b)$ we obtained:

$$\begin{pmatrix} \left(b_1^m M^2 + g_{11}P_3^2 M^2\right) 2\sin(\theta_a - \theta_b)\cos(\theta_a + \theta_b) + \\ + \left(c_1^m M^2 + f_{11}P_3^2 M^2\right)\sin(\theta_a - \theta_b) + \delta\dfrac{d}{dz}\left[2M^2\left(\dfrac{d(\theta_a - \theta_b)}{dz}\right)\right] \end{pmatrix} = 0 \quad \text{(A.27a)}$$

$$\begin{pmatrix} -2c(T)M^2\sin(\theta_a + \theta_b) - \left(b_1^m M^2 + g_{11}P_3^2 M^2\right)2\sin(\theta_a + \theta_b)\cos(\theta_a - \theta_b) - \\ -\left(c_1^m M^2 + f_{11}P_3^2 M^2\right)\sin(\theta_a + \theta_b) - \delta\dfrac{d}{dz}\left[2M^2\left(\dfrac{d(\theta_a + \theta_b)}{dz}\right)\right] - 2M H_0 \end{pmatrix} = 0 \quad \text{(A.27b)}$$

In the absence of magnetic field $H_0=0$ and neglecting gradients, Eqs.(A.27) reduce to

$$\left(b_1^m + g_{11}P_3^2\right)2\sin(\theta_a - \theta_b)\cos(\theta_a + \theta_b) + \left(c_1^m + f_{11}P_3^2\right)\sin(\theta_a - \theta_b) = 0 \quad \text{(A.28a)}$$

$$2c(T)M\sin(\theta_a + \theta_b) + M\left(b_1^m + g_{11}P_3^2\right)2\sin(\theta_a + \theta_b)\cos(\theta_a - \theta_b) + \\ + M\left(c_1^m + f_{11}P_3^2\right)\sin(\theta_a + \theta_b) + 2H_0 = 0 \quad \text{(A.28b)}$$

If $H_0 = 0$ and $\theta_a - \theta_b \neq \pi n$, $n \in N$, then one can reduce the system to the following:

$$\cos(\theta_a + \theta_b) = -\dfrac{c_1^m + f_{11}P_3^2}{\left(b_1^m + g_{11}P_3^2\right)2} \text{ and } \cos(\theta_a - \theta_b) = -\dfrac{2c(T) + c_1^m + f_{11}P_3^2}{2\left(b_1^m + g_{11}P_3^2\right)} \ll -1, \text{ since } c(T) \gg b_1^m$$

for typical situations (i.e. $\theta_a - \theta_b \neq \pi n$, $n \in N$ has no sense). Thus, if $H_0 = 0$, only the relations $\theta_a + \theta_b = \pi n$ or $\theta_a - \theta_b = \pi m$ have sense. In other words, there exist four states, namely:

1) weak-axis antiferromagnetic with $\theta_a = 0$, $\theta_b = \pi$ or vise versa;

2) weak-plain antiferromagnetic with $\theta_a = \theta_b = \dfrac{\pi}{2}$ (or $\theta_a = \theta_b = 3\dfrac{\pi}{2}$);

3) weak-axis ferromagnetic with $\theta_a = 0$, $\theta_b = 0$ (or $\theta_a = \pi$, $\theta_b = \pi$);

4) weak-plain ferromagnetic with $\theta_a = \dfrac{\pi}{2}$, $\theta_b = 3\dfrac{\pi}{2}$ (or $\theta_a = 3\dfrac{\pi}{2}$, $\theta_b = \dfrac{\pi}{2}$).

They could be either stable phases (minima) or unstable states (maxima, separating different minima).

**Appendix B. Parameters estimations**

In Curie-Weiss model (or mean field approximation for ferromagnetic) local magnetic field, acting on magnetic moments, is approximated as $H_{eff}=H_0+\nu M$, where $\nu$ is Weiss constant of molecular field. Therefore, magnetization has to be found from the following equation (see e.g. pp. 101-102 in Ref. [24]):

$$M = N_m \mu_B \tanh\left(\dfrac{\mu_B \mu_0 (H_0 + \nu M)}{k_B T}\right) \quad \text{(B.1)}$$

Here $N_m$ is the concentration of magnetic moments, $\mu_B \approx 0.927 \cdot 10^{-23}$ A·m² is Bohr magneton, $\mu_0 \approx 4 \cdot \pi 10^{-7}$ N/A² is magnetic constant, $k_B = 1.38 \cdot 10^{-23}$ J/K is Boltzmann constant. Using typical concentration range $N_m \sim 10^{28} \div 10^{29}$ m⁻³, one can see from Eq. (B.1) that maximal value of $M$ is $N_m \mu_B$ which is about $10^5 \div 10^6$ A/m (corresponding magnetic induction of saturation, $N\mu_B\mu_0$, is about $0.1 \div 1$ Tesla; for comparison, for Fe it is about 2 Tesla).

In paramagnetic phase, for very small magnetic field $H_0 \to 0$, $M \to 0$, one can find from (B.1):

$$M = \frac{N_m \mu_B^2 \mu_0}{k_B T - N_m \mu_B^2 \mu_0 \nu} H_0 \tag{B.2}$$

It is seen from (B.2), that Curie (or Neel) temperature is $T_{C,N} = N_m \mu_B^2 \mu_0 \nu / k_B$, while Curie-Weiss constant is $C_{CW} = N_m \mu_B^2 \mu_0 / k_B$. Using typical values, one can easily find that $C_{CW} \sim 10^{-1} \div 10^0$ K, while for using values $T_{C,N} \sim 10^2 \div 10^3$ K, it is obvious that $\nu \sim 10^2 \div 10^4 \gg 1$.

Using the phenomenological expansion of free energy

$$\alpha_M (T - T_N) \mathbf{M}^2 + d\, \mathbf{M}^4 - \mu_0 \mathbf{M}\, \mathbf{H}_0 \tag{B.3}$$

one can easily find the following relations, in paramagnetic phase $\mathbf{M}(T > T_N) = \dfrac{\mu_0 \mathbf{H}_0}{2\alpha_M (T - T_N)}$, while in ferromagnetic phase spontaneous magnetization is $\mathbf{M}(T \to 0) = \sqrt{\dfrac{\alpha_M T_N}{2d}}$. Comparing these relations with (B.1) and (B.2), one can easily estimate that

$$\alpha_M = \frac{k_B}{2 N_m \mu_B^2} \sim 10^{-5} \div 10^{-6}\ \frac{J}{K\, m\, A^2} \quad \text{and} \quad d = \frac{\alpha_M T_N}{2 N_m^2 \mu_B^2} = \frac{\mu_0 \nu}{4 N_m^2 \mu_B^2} \sim 10^{-13} \div 10^{-17}\ \frac{J\, m}{A^4}.$$

Magnetostriction coefficients estimation can be done using phenomenological relation for magnetostriction strain $u_M = Z \cdot M^2$. Using typical values $u_M \sim 10^{-4} \div 10^{-6}$ and $M \sim 10^5 \div 10^6$ A/m, one can obtain $Z \sim 10^{-14} \div 10^{-18}$ m²/A².

Exchange integral is of order $T_{C,N} k_B = 10^{-21} \div 10^{-20}$ J. Using values $N_m \sim 10^{28} \div 10^{29}$ m⁻³, $\mu_B \approx 0.927 \cdot 10^{-23}$ A·m² we estimate that $c = T_{C,N} k_B / N_m (\mu_B)^2 = 10^{-4} \div 10^{-2}$ J/(m A²)

When generating phase diagrams we used the following range of magnetic parameters: magnetostriction coefficients $W_{ij} = 10^{-14} \div 10^{-18}$ m²/A², $Z_{ij} = 10^{-14} \div 10^{-18}$ m²/A² and we assume that they obey the same interrelations that electrostriction coefficients

$\alpha_M = 10^{-5} \div 10^{-6}$ J/(K m A²), $T_{C,N} = 10^2 \div 10^3$ K, $\delta = 10^{-20}$ m J A⁻²,

$d = 10^{-13} \div 10^{-17}$ J m A⁻⁴, $k = 10^{-15} \div 10^{-19}$ J m A⁻⁴,

$c=10^{-5} \div 10^{-2}$ J/(m A$^2$), $b_1=10^{-9} \div 10^{-4}$ J/(m A$^2$), $c_1=10^{-9} \div 10^{-4}$ J/(m A$^2$)

Misfit strain $u_m \sim 10^{-2} \div 10^{-3}$.

Finally, let us estimate the coefficient $A_{11}$. This could be done from the jump of elastic compliance $s_{11}$ in the point of the bulk ferroelectric phase transition, since $\Delta s_{11}^{P,E} = \left(s_{11}^{P,E} - s_{11}\right) \sim P_3^2$, namely $\Delta s_{12}^{P,E} = A_{11} P_3^2$. In accordance with data for BaTiO$_3$ of Ref.[27] one obtains that $\Delta s_{11}^P = -1.6 \cdot 10^{-12}$ Pa and $\Delta s_{11}^E = 6 \cdot 10^{-12}$ Pa; whereas $P_3^2 = 0.2$ C/m$^2$ at 100$^0$C. Thus $A_{11}^P = -4 \cdot 10^{-11}$ m$^4$/C$^2$ Pa and $A_{11}^E = +15 \cdot 10^{-11}$ m$^4$/C$^2$Pa.

**References**

1. J.F. Scott. Nature Materials, **6**, 256 (2007)

2. E. Roduner. *Nanoscopic Materials. Size-dependent phenomena*. RSC Publishing (2006).

3 M. Fiebig, J. Phys. D: Appl. Phys. **38**, R123-R152 (2005).

4. V.K. Wadhawan. *Introduction to ferroic materials*. Gordon and Breach Science Publishers (2000).

5 M.G. Cottan. *Linear and nonlinear spin waves in magnetic films and super-lattices*. World Scientific, Singapore (1994).

6 Ce Wen Nan, Phys. Rev. B **50**, 6082 (1994).

7 Ce Wen Nan, Ming Li, Jin H. Huang, Phys. Rev. B **63**, 144415 (2001).

8 S. Dong, J.F. Li, D. Viehland, Phyl. Magazine, **83**, p.769-773 (2003).

9 J.X. Zhang, Y.L. Li, D.G. Schlom, L.Q. Chen, F. Zavaliche, R. Ramesh, and Q.X. Jia. Appl. Phys. Lett. **90**, 052909 (2007).

10 J. Hemberger, P. Lunkenheimer, R. Fichtl, H.-A. Krug von Nidda, V. Tsurkan, and A. Loidl, Nature **434**, 364 (2005).

11 M.D. Glinchuk, E.A. Eliseev, A.N. Morozovska, and R.Blinc, Phys. Rev. B **77**, 024106 (2008).

12 J. Wang, J.B. Neaton, H. Zheng, V. Nagarajan, S.B. Ogale, B. Liu, D. Viehland, V. Vaithyanathan, D.G. Schlom, U.V. Waghmare, N.A. Spaldin, K.M. Rabe, M. Wuttig, and R. Ramesh, Science **299**, 1719 (2003).

13 W. Tian, V. Vaithyanathan, D.G. Schlom, Q. Zhan, S.Y. Yang, Y.H. Chu, and R. Ramesh, Appl. Phys. Lett. **90**, 172908 (2007).

14 H. Naganuma, N. Shimura, J. Miura, H. Shima, Sh. Yasui, K. Nishida, T. Katoda, T. Iijima, H. Funakubo, and S. Okamura, J. Appl. Phys. **103**, 07E314 (2008).

15 B. Ruette, S. Zvyagin, A.P. Pyatakov, A. Bush, J.F. Li, V.I. Belotelov, A.K. Zvezdin, and D. Viehland, Phys. Rev. B **69**, 064114 (2004).

16 M.I. Kaganov and A.N. Omelyanchouk, Zh. Eksp. Teor. Fiz. **61**, 1679 (1971) [Sov. Phys. JETP **34**, 895 (1972)].

17 R. Kretschmer, K. Binder, Phys. Rev. B **20**(3), 1065-1076 (1976).

18 J.S. Speck and W. Pompe, J. Appl. Phys. **76**, 466 (1994).

19 M.D. Glinchuk, A.N. Morozovska, J. Phys.: Condens. Matter. **16**, 3517 (2004).

20 C.-L. Jia, V. Nagarajan, J.-Q. He, L. Houben, T. Zhao, R. Ramesh, K. Urban, and R. Waser. Nature Mat. **6**, 64 (2007).

21 V.V. Eremenko, and V.A. Sirenko, *Magnetic and magneto-elastic properties of atniferromagnets and superconductors* (Naukova Dumka, Kiev, 2004) in rus.

22 M.D. Glinchuk, A.N. Morozovska, and E.A. Eliseev, J. Appl. Phys, **99**, 114102 (2006)

23 see Supplement in ArXiv: A.N. Morozovska, M.D. Glinchuk, E.A. Eliseev, and R. Blinc, Arxiv.org/abs/0803.4246

24 M. I. Kaganov, and V.M. Tsukernik, *Nature of Magnetism*, (Nauka, Moscow, 1982) in rus.

25 L.D. Landau, E.M. Lifshitz, and L. P. Pitaevskii. Electrodynamics of Continuous Media, (Second Edition Butterworth-Heinemann, Oxford, 1984).

26 J.X. Zhang, Y.L. Li, Y. Wang, Z.K. Liu, L.Q. Chen, Y.H. Chu, F. Zavaliche, and R. Ramesh, J. Appl. Phys. **101**, 114105 (2007).

27 F. Jona and G. Shirane, *Ferroelectric Crystals* (Dover Publications, New York, 1993). p.208